\documentclass[aps,prl,twocolumn,amssymb,nofootinbib]{revtex4}
\usepackage{graphicx}
\usepackage{amsmath}
\usepackage{amssymb}
\usepackage{bm}

\usepackage{epsfig}
\usepackage{bbm}
\usepackage{color}
\usepackage{ulem}

\definecolor{MyDarkBlue}{rgb}{0.15,0.15,0.45}
\usepackage[linktocpage=true]{hyperref}
\hypersetup{
colorlinks=true,
citecolor=MyDarkBlue,
linkcolor=MyDarkBlue,
urlcolor=MyDarkBlue,
pdfauthor={Garrett Goon, Kurt Hinterbichler, Austin Joyce and Mark Trodden},
pdftitle={Gauged Galileons From Branes},
pdfsubject={hep-th}
}

\usepackage{enumitem}
\usepackage{graphicx}
\usepackage{subfigure}
\usepackage{amsmath}
\usepackage{bbm}
\usepackage{color}

\newcommand{\be}{\begin{equation}}
\newcommand{\ee}{\end{equation}}
\newcommand{\bea}{\begin{eqnarray}}
\newcommand{\eea}{\end{eqnarray}}
\newcommand{\beas}{\begin{eqnarray*}}
\newcommand{\eeas}{\end{eqnarray*}}
\newcommand{\nn}{\nonumber\\}

\def\({\left(}
\def\){\right)}

\begin{document}

\title{Gauged Galileons From Branes} 
\author{Garrett L. Goon}
\author{Kurt Hinterbichler}
\author{Austin Joyce}
\author{Mark Trodden}

\affiliation{Center for Particle Cosmology, Department of Physics and Astronomy, University of Pennsylvania,
Philadelphia, Pennsylvania 19104, USA}
\date{\today}

\begin{abstract}
We show how the coupling of $SO(N)$ gauge fields to galileons arises from a probe brane construction.  The galileons arise from the brane bending modes of a brane probing a co-dimension $N$ bulk, and the gauge fields arise by turning on certain off-diagonal components in the zero mode of the bulk metric.  By construction, the equations of motion for both the galileons and gauge fields remain second order.  Covariant gauged galileons are derived as well.
\end{abstract}

\maketitle
\section*{Introduction}
Many recent investigations have involved---either directly or indirectly---the presence of galileons, which are higher-derivative scalar fields that both have second-order equations of motion and are also invariant under a novel ``galilean" symmetry: $\pi(x)\to\pi(x)+c+b_\mu x^\mu$. Originally, this symmetry arose in the scalar sector of the decoupling limit of the Dvali--Gabadadze--Porrati (DGP) model \cite{Dvali:2000hr}, where it may be thought of as a small-field consequence of the nonlinearly-realized five-dimensional Poincar\'e symmetry.  This galilean symmetry has since been abstracted and studied in its own right \cite{Nicolis:2008in} (for a review of recent developments, see \cite{Trodden:2011xh}). Galileons have rather nice properties and structure; it is non-trivial that there exist terms invariant under the galilean symmetry which also have second order equations of motion for the field $\pi$.  Second order equations of motion guarantee that the theory does not propagate extra ghostly degrees of freedom which are common in other higher-derivative theories.  Further, choosing to consider only these terms is consistent from an effective field theory viewpoint; the fact that they have fewer derivatives than other terms invariant under the galilean shift symmetry means that there exists a regime where galileons are the dominant terms and the others can be consistently neglected \cite{Nicolis:2004qq,Endlich:2010zj,Hinterbichler:2010xn}. Additionally, due to their symmetry properties and the fact that they shift by a non-trivial total derivative under the symmetry (they are Wess-Zumino terms \cite{Goon:2012dy}) , galileon theories are radiatively stable---they are not renormalized at any loop order in perturbation theory \cite{Luty:2003vm, Hinterbichler:2010xn}. 

The properties of galileons are simple to state, but the theories possess a rich and interesting phenomenology. Galileons have been used to address issues in both the early universe through inflation \cite{Kobayashi:2010cm, Burrage:2010cu, Creminelli:2010qf} and alternatives to inflation \cite{Creminelli:2010ba}, as well as in the late universe where they have been investigated as a possible source of cosmic acceleration \cite{Chow:2009fm, Silva:2009km, DeFelice:2010as, Deffayet:2010qz}. Galileons also make an appearance in ghost-free massive gravity, where they describe the interactions of the longitudinal polarization of the graviton in the decoupling limit \cite{deRham:2010ik,deRham:2010kj} (for a review see \cite{Hinterbichler:2011tt}).  

Many applications require that galileon theories be covariantized. This is possible, but retaining their second-order equations of motion requires introducing non-minimal coupling between the fields and curvature, generically destroying the shift symmetry of the field \cite{Deffayet:2009wt, Deffayet:2009mn, Deffayet:2011gz}.  Appropriate non-minimal terms arise naturally in the probe brane construction~\cite{deRham:2010eu}; this construction also elucidates the origin of the second-order equations of motion---the galileon terms descend from Lovelock invariants of the induced brane metric and from Gibbons--Hawking--York (GHY) boundary terms associated to bulk Lovelock invariants. The Lovelock terms are of course the only terms that may be added to Einstein gravity while maintaining second order metric equations of motion \cite{Lovelock:1971yv}, and this property is passed down to the galileons through the probe brane construction.

The probe brane construction has been extended to curved brane backgrounds, on which fields are invariant under intricate nonlinear symmetries inherited from the isometries of the bulk \cite{Goon:2011qf,Goon:2011uw, Burrage:2011bt,Goon:2011xf}. The brane construction has also been generalized to higher co-dimension \cite{Hinterbichler:2010xn}; this generalization leads to a multi-galileon theory where the fields possess an internal global $SO(N)$ symmetry, which is inherited from the symmetries of the higher co-dimension bulk. Related multi-galileon theories were discussed in \cite{Deffayet:2010zh,Padilla:2010de,Padilla:2010ir,Padilla:2010tj}.

Recently it was shown by Zhou and Copeland \cite{Zhou:2011ix} that it is possible to couple galileons to gauge fields while retaining second-order equations of motion. In this note we generalize the higher dimensional probe brane construction of \cite{Hinterbichler:2010xn} to recover the $SO(N)$ gauged galileon theories of \cite{Zhou:2011ix} from a purely geometric setup.

\section*{Gauging the Galileons}
An $N$-galileon theory contains $N$ scalar fields $\pi^I$, indexed by $I=1,\cdots,N$, which have second order equations of motion and a galilean and shift symmetry on each field: $\pi^I(x)\to\pi^I(x)+c^I+b^I_\mu x^\mu$, where $b^I_{\mu}$ and $c^I$ are constants.    There may also exist global internal symmetries under which $\pi^I$ transforms in a linear representation \cite{Padilla:2010ir,Hinterbichler:2010xn}.  These global symmetries can be promoted to local ones \cite{Zhou:2011ix}.  Here we will restrict to the case where the galileons transform as a fundamental under $SO(N)$, the case which naturally follows from a co-dimension $N$ brane construction \cite{ Hinterbichler:2010xn}, since our goal here is to demonstrate a brane perspective for deriving these gauged multi-galileons.

On flat space, there are two multi-galileon Lagrangians which respect the $SO(N)$ global internal symmetry for $N\ge 2$ \cite{Padilla:2010ir,Hinterbichler:2010xn}, 
\begin{align}
\mathcal{L}_{2}&= -{1\over 2}\partial_{\mu}\pi^{I}\partial^{\mu}\pi_{I},\nn
\mathcal{L}_{4}&=-\partial_\mu\pi^I\partial_\nu\pi^J\left(\partial_\lambda\partial^\mu\pi_J\,\partial^\lambda\partial^\nu\pi_I-\partial^\mu\partial^\nu\pi_J\,\square\pi_I \right)\label{allflatNgalileons}~.
\end{align}
The symmetries of these terms come in three sets \cite{Hinterbichler:2010xn}
\begin{align}
\delta_{1}\pi^{I}&=-\omega^{\mu}{}_{\nu}x^{\nu}\partial_{\mu}\pi^{I}-\epsilon^{\mu}\partial_{\mu}\pi^{I},\nn
\delta_{2}\pi^{I}&=b^{I}_{\mu}x^{\mu}+c^{I},\\
\nonumber
\delta_{3}\pi^{I}&=\omega^{I}{}_{J}\pi^{J}.\
\end{align}
The first is ordinary Poincar\'e invariance for the scalar fields, the second is the galilean and shift symmetry, and the third is the internal $SO(N)$ symmetry (for which $\omega_{IJ}$ is the infinitesimal antisymmetric parameter). 

As considered in \cite{Zhou:2011ix}, we may promote the global $SO(N)$ symmetry to a local one by minimal substitution, 
$\partial_{\mu}\pi^I\to D_{\mu}\pi^I=\partial_{\mu}\pi^I+A_{\mu J}^I\pi^J$.
Here $A_{\mu J}^I$ is an anti-symmetric matrix, which is just the gauge connection in the fundamental representation of $SO(N)$,
\be A_{\mu J}^I=-{i\over 2}A_\mu^{KL}\(T_{KL}\)^I_{\ J}, \ee
where the generators of $SO(N)$ are given by 
\be \(T_{KL}\)^I_{\ J}=i\(\delta_K^I\delta_{LJ}-\delta_L^I\delta_{KJ}\). \ee
This minimal coupling procedure gives gauge invariant actions with second order equations of motion both for $\pi^{I}$ and $A_{\mu J}^I$.  The gauging, however, eliminates the galilean symmetry (this is similar to the situation that occurs when covariantizing the galileons).

The presence of second order equations of motion after the na\"ive gauging is not surprising, as was pointed out in \cite{Zhou:2011ix}.  Due to the structure of the spacetime index contractions, there will never be more than two derivatives on a $\pi^{I}$ and the highest derivatives on $A_{\mu}$ enter through expressions of the form $D_{\lambda}F_{\mu\nu}$, where $F_{\mu\nu}$ is the field strength
\be F_{\mu\nu}=\partial_\mu A_\nu-\partial_\nu
A_\mu+\left[A_\mu,A_\nu\right].\ee
Since $F_{\mu\nu}$ contains only first derivatives on $A_{\mu}$, the equations of motion for $A_{\mu}$ are at most second order. 

Note that the minimal coupling prescription is ambiguous.  For instance, one could have changed the ordering of derivatives in the action, say $\partial_{\mu}\partial_{\nu}\pi^{I}\to \partial_{\nu}\partial_{\mu}\pi^{I}$, and gauging the Lagrangians before and after this replacement would give different results, since the gauge covariant derivatives do not commute,
\be [D_\mu,D_\nu]\pi^I={F_{\mu\nu}}^I_{\ J}\pi^J.\ee   We can't say one choice is ``more minimal''  than the other.  Requiring second order equations of motion does not pin down the Lagrangian uniquely, since there is freedom to add non-minimal terms (even beyond those resulting from commuting derivatives) which do not lead to higher order equations.  One of the virtues of the brane construction will be to pick out a particular set of non-minimal couplings.

We may also consider coupling to gravity through na\"ive covariantization, $\partial_{\mu}\to\nabla_{\mu}$.  This maintains second order equations of motion for $\pi^{I}$, but there also arise terms of the form $\nabla_{\lambda}R_{\mu\nu\rho\sigma}$ in the equation of motion of $\mathcal{L}_{4}$.  As $R$ is second order in derivatives of the metric, the equation of motion is third order in the metric.  Adding a non-minimal coupling can remove these third order derivatives and those in the metric equations of motion \cite{Deffayet:2009wt}; for example the following has second order equations of motion for both the scalars and metric,
\begin{align}
\mathcal{L}_{4,{\rm cov}}&= -\nabla_{\mu}\pi^{I}\nabla_{\nu}\pi^{J}\Big(\nabla^{\nu}\nabla^{\lambda}\pi_{I} \nabla^{\mu}\nabla_{\lambda}\pi_{J}-\nabla^2\pi_{I}\nabla^{\mu}\nabla^{\nu}\pi_{J}\Big)\nn
&\quad -\left (R_{\mu\nu}-\frac{1}{4}Rg_{\mu\nu}\right )\Big(\nabla^{\mu}\pi^{I}\nabla^{\nu}\pi^{J}\nabla^{\lambda}\pi_{I}\nabla_{\lambda}\pi_{J}\nn
&\quad -\frac{1}{2}\nabla^{\mu}\pi^{I}\nabla^{\nu}\pi_{I}\nabla^{\lambda}\pi^{J}\nabla_{\lambda}\pi_{J}\Big)\ .\label{covariantizedR}
\end{align}

This lagrangian can also be obtained from the probe brane construction \cite{Hinterbichler:2010xn}. As in the case of gauging, and as seen from \cite{Deffayet:2011gz}, the choice of non-minimal terms in (\ref{covariantizedR}) is not unique; there are other possible non-minimal couplings which still give second order equations of motion for all the fields.  No choice is singled out by the procedure of minimal coupling followed by the addition of non-minimal terms to cancel higher-order pieces of the equations of motion, and a virtue of the brane construction will be to single out a specific choice of non-minimal terms.

Covariantizing the galileons in this way breaks the galilean symmetry, but preserves the global $SO(N)$, which can then be gauged by replacing $\nabla_{\mu}\to \mathcal{D}_{\mu}=\nabla_{\mu}+A_{\mu}$.  The resulting gauge and diffeomorphism invariant Lagrangian has second order equations of motion for the scalars, the metric, and the gauge fields \cite{Zhou:2011ix}.

\section*{The Higher Dimensional Brane Construction}

In this section, we briefly review the probe brane construction of the multi-galileons.  For a more detailed treatment, we refer the reader to \cite{Hinterbichler:2010xn,Goon:2011qf}.

The probe brane construction was originally developed~\cite{deRham:2010eu} for single field galileons arising via a co-dimension one brane probing a flat bulk.  The action is constructed from diffeomorphism scalars formed from the induced metric and extrinsic curvature of a 3-brane floating in the 5D bulk.  Symmetries of the action are inherited from Killing vectors of the bulk \cite{Goon:2011qf} and the unique co-dimension one Lagrangians which have second order equations of motion are the 4D Lovelock invariants and the Gibbons--Hawking--York boundary terms for the 5D Lovelock invariants (and a tadpole term).  

Extending the probe brane construction to higher co-dimension allows for the construction of multi-galileon theories \cite{Hinterbichler:2010xn}.
We begin with a $D$-dimensional bulk with coordinates $X^{A}$ and metric $G_{AB}(X)$.  The position of a $4$-dimensional brane living in the bulk is given by embedding functions $X^{A}(x)$, where $x^{\mu}$ are coordinates on the brane.  Tangent vectors to the brane have components $e^{A}_{\mu}=\frac{\partial X^{A}}{\partial x^{\mu}}$ and the induced metric on the brane is
\begin{align}
\bar{g}_{\mu\nu}&=e^{A}_{\mu}e^{B}_{\nu}G_{AB}\ .
\end{align}
There are also $N\equiv(D-4)$ vectors normal to the brane indexed by $I$, with components $n^A_I$, which satisfy
\begin{align}
n^{A}_Ie^{B}_{\mu}G_{AB}=0,\ \ \ n^{A}_In^{B}_JG_{AB}=\delta_{IJ}.\
\end{align}
The normal and tangent vectors are used to construct the $N$ extrinsic curvature tensors,
\begin{align}
K^{I}_{\mu\nu}&=e^{A}_{\mu}e^{B}_{\nu}\nabla_{A}n^{I}_{B}\ ,
\end{align}
where $\nabla_{A}$ is the bulk covariant derivative, as well as the twist connection, which is the connection on the normal bundle, 
\be\beta_{\mu J}^I=n^{B I}e^A_{\ \mu}\nabla_A n_{BJ}~;\ee
it has an associated curvature $R^I_{~J\mu\nu}$.

Requiring the action to be invariant under reparametrizations of the brane restricts the action to be a diffeomorphism scalar constructed from these geometric ingredients,
\begin{align}
S&=\int {\rm d}^{4}x\, \sqrt{-g}\,\mathcal{L}(\bar{g}_{\mu\nu},\bar{\nabla}_{\mu}, \bar{R}^\mu_{~\nu\rho\sigma},K^{I}_{\mu\nu}, R^I_{~J\mu\nu})\ .\label{genericaction}
\end{align}
Here $\bar{\nabla}_{\mu}$ is the world-volume connection, which acts on 4D spacetime indices with the Levi--Civita connection of the induced metric, and on normal indices with the twist connection.  
We fix the reparametrization symmetry of the brane worldvolume coordinates by choosing
\begin{align}
X^{\mu}(x)=x^{\mu}, \ \ \ X^{I}(x)=\pi^{I}(x),\ \ \label{gaugechoice}
\end{align}
that is, we take the $4$ worldvolume coordinates to coincide with the first 4 coordinates used in the bulk.  The $N$ remaining functions $\pi^{I}$ are the physical degrees of freedom for the brane.

Given a Killing vector $K^{A}$ of the bulk metric $G_{AB}$, the induced metric and extrinsic curvature (and hence the action (\ref{genericaction})) are invariant under $\delta_{K} X^{A}=K^{A}$.  However, generically this destroys the gauge choice (\ref{gaugechoice}) by sending
\begin{align}
x^{\mu}\to x^{\mu}+K^{\mu},
\end{align}
and we must restore the desired gauge via a brane reparametrization $\delta_{g} X^{A}(x)=\xi^{\mu}\partial_{\mu}X^{A}(x)$ with $\xi^{\mu}=-K^{\mu}$ so that the combined gauge-preserving $\pi^{I}$ symmetry acts as
\begin{align}
(\delta_{K}+\delta_{g})\pi^{I}=-K^{\mu}\partial_{\mu}\pi^{I}+K^{I} ,
\end{align}
and becomes a global symmetry of the gauge fixed action.
Symmetries that have a $K^I$ component are nonlinearly realized and are thus symmetries of the bulk that are spontaneously broken due to the presence of the brane.

Generic choices of the action (\ref{genericaction}) will not give second order equations of motion for the $\pi^{I}$. For a $4$-dimensional brane the unique terms that give second-order equations of motion are the $4$-dimensional Lovelock terms and possible Gibbons--Hawking--York boundary terms for the higher dimensional Lovelock terms, whose specific form depends on the dimensions of the brane and the number of co-dimensions.  A four dimensional brane has two 4D Lovelock terms---the cosmological constant and the induced Ricci curvature,
\begin{align}
\mathcal{L}_{2}&=-\sqrt{-\bar g}\nn
\mathcal{L}_{4}&=-\sqrt{-\bar g}\,\bar{R}\ .\label{twolovelocks}
\end{align}
The possible GHY terms for the 3-brane depends on the number of co-dimensions \cite{Charmousis:2005ey, Charmousis:2005ez, Hinterbichler:2010xn}. However, as was shown in \cite{Hinterbichler:2010xn}, in the end no new possibilities for actions are generated beyond those given by (\ref{twolovelocks}), so we need only consider these two.

The galileons are obtained by taking the bulk metric to be fixed and flat, $G_{AB}(X)=\eta_{AB}$.  The induced metric is $\bar g_{\mu\nu}=\eta_{\mu\nu}+\partial_\mu\pi^I\partial_\nu\pi_I$.  Evaluating the actions (\ref{twolovelocks}) gives relativistic DBI versions of the $SO(N)$ symmetric galileons.  A small field limit then reproduces (\ref{allflatNgalileons}).  The flat metric has maximal symmetry and all of these symmetries are realized in the galileon theory.  The Poincar\'e transformations along the brane become the 4D Poincar\'e transformations, the rotations in the extra dimensions become the internal $SO(N)$ symmetry, translations in the extra dimensions become the shift symmetry, and the (small field limit of) boosts into the extra dimensions become the galilean symmetry. The small field limit may be viewed as either an expansion in derivatives or in fields, both produce the same result \cite{deRham:2010eu,Goon:2011qf}. From an algebraic perspective, the small field limit may be thought of as Wigner--\.In\"on\"u contraction of the Poincare algebra along the co-dimension directions, that is, sending the speed of light in the directions away from the brane to infinity \cite{Goon:2012dy}.

\section*{Gauged Galileons from Branes}

We now show how to obtain gauged symmetries from the previously discussed probe brane description. To gauge the symmetries, we simply turn on zero modes for the background metric $G_{AB}(X)$.  

For example, to couple to gravity, we take the background metric to be \cite{deRham:2010eu}
\begin{align}G_{AB}(X)&=\left (
\begin{array}{c|c}
g_{\mu\nu}(x)&0 \\  \hline
0 & \delta_{ab}
\end{array}\right )\ .
\label{bulkmetricg}
\end{align}
We have turned on the 4D part of the metric and allowed it to depend only on the 4D coordinates $x^\mu$.  The induced metric now becomes $\bar g_{\mu\nu}=
g_{\mu\nu}+\nabla_\mu\pi^I\nabla_\nu\pi_I$.  Evaluating the actions (\ref{twolovelocks}) gives relativistic DBI versions of the covariant $SO(N)$ symmetric galileons.  A small field limit then reproduces precisely (\ref{covariantizedR}) and the canonical kinetic term \cite{Hinterbichler:2010xn}.   The non-minimal terms in ${\cal L}_{4,{\rm cov}}$ needed to make the equations of motion second order come out automatically, and a unique such term is produced. 

The metric (\ref{bulkmetricg}) breaks the higher-dimensional Poincar\'e invariance.  All that survives is the $SO(N)$ rotations and translations in the extra dimensions.  This is reflected in the fact that the only symmetries left in (\ref{covariantizedR}) are $SO(N)$ rotations and shifts on the fields.  The extended galilean symmetry is lost.  The zero mode metric $g_{\mu\nu}$ and the scalars $\pi^I$ inherit a diffeomorphism transformation under the zero mode of higher-dimensional diffeomorphisms which preserves the ansatz (\ref{bulkmetricg}), and this yields the diffeomorphism invariance of the 4D theory.

To gauge the $SO(N)$ internal symmetry, we will turn on zero modes of off-diagonal components of the background metric, corresponding to Killing vectors of the extra dimensions.  We take a bulk metric of the form seen in Kaluza--Klein reductions
\begin{align}G_{AB}&=\left (
\begin{array}{c|c}
\eta_{\mu\nu}+A_{\mu}^{i}(x)A_{\nu}^{j}(x)\xi^{I}_i(y)\xi_{jI}(y)& A^{i}_{\mu}(x)\xi_{iI}(y)\\ \hline
A^{i}_{\mu}(x)\xi_{iI} (y)& \delta_{IJ}
\end{array}\right )\ .
\label{gaugeonlyans}
\end{align}
The $\xi^{I}_i(y)$'s are Killing vectors of $\delta_{IJ}$ (depending on $y^I$, the coordinates in the extra dimensions), $I$ denotes the components of the Killing vector in the extra-dimensional space and $i$ labels the various Killing vectors.  The coefficient functions $A^{i}_{\mu}(x)$ are arbitrary functions of the 4D coordinates which will be the gauge fields from the perspective of the brane.  
The induced metric on a $4$-dimensional brane, calculated in the gauge (\ref{gaugechoice}), is now given by
\begin{align}
\bar{g}_{\mu\nu}&=\eta_{\mu\nu}+\left (\partial_{\mu}\pi^{I}+A_{\mu}^{i}\xi^{I}_i(\pi)\right )\left (\partial_{\nu}\pi_I+A_{\nu}^{j}\xi_{jI}(\pi)\right )\ .
\end{align}

We want to gauge only $SO(N)$, so we will turn on only those Killing vectors corresponding to rotations in the extra dimensions.\footnote{Including the translational Killing vectors would result in gauging the shift symmetry of the galileons.  The galileons would then be pure gauge, and would become longitudinal components of the translational gauge fields.}  The index $i$ can then be taken to be the anti-symmetric index set $[JK]$ which runs over $N(N-1)/2$ values.  The components of the Killing vectors $\xi^{[JK]I}$ are given by
\begin{align}
\xi^{[JK]I}(y)=y^{K}\delta^{JI}-y^{J}\delta^{KI}.
\end{align}

Now we have $\partial_{\mu}\pi^{I}+A_{\mu}^{i}\xi^{I}_i(\pi)=\partial_{\mu}\pi^{I}+{1\over 2}A_{\mu}^{JK}\xi^{I}_{[JK]}(\pi)=\partial_{\mu}\pi^{I}+A_{\mu J}^I\pi^J=D_\mu\pi^I$,
 and we recover the covariant derivatives, so the induced metric indeed takes the form conjectured in \cite{Zhou:2011ix}, 
\begin{align}\label{gaugeonlycov}
\bar{g}_{\mu\nu}&=\eta_{\mu\nu}+{D}_{\mu}\pi^{I}{D}_{\nu}\pi_{I}\ .
\end{align}

Evaluating the action (\ref{twolovelocks}) now gives a relativistic DBI version of the gauged $SO(N)$ galileons, whose small field limits reproduce the gauged galileons of \cite{Zhou:2011ix}. For example, the  gauged kinetic term comes from the cosmological term $-\sqrt{-\bar g}$ and expanding to quadratic order in $\pi$, we find
\begin{align}
\nonumber
\mathcal{L}_{2,{\rm gauged}}&=-\sqrt{-\bar{g}}=-\sqrt{-\det\left(\eta_{\mu\nu}+{D}_{\mu}\pi^{I}{D}_{\nu}\pi_{I}\right)}\\ 
&= -1-\frac{1}{2}{D}_{\mu}\pi^{I}{D}^{\mu}\pi_{I} +{\cal O}\left(\pi^4\right).\,
\end{align}
The Einstein-Hilbert term yields, in the small field limit,
\begin{align}
\nonumber
&\mathcal{L}_{4,{\rm gauged}} = -\sqrt{-\bar{g}}\bar R\rightarrow\\
\nonumber
&\quad -{D}_{\mu}\pi^{I}{D}_{\nu}\pi^{J}\Big({D}^{\nu}{D}^{\lambda}\pi_{I} {D}^{\mu}{D}_{\lambda}\pi_{J}-{D}^2\pi_{I}{D}^{\mu}{D}^{\nu}\pi_{J}\Big)\\
\nonumber
&\quad +{1\over 2}\(F^{\mu\nu}\pi\)^{I}{D}_{\lambda}\pi^{J}{D}^{\lambda}\pi_{I}{D}_{\mu}{D}_{\nu}\pi_{J}\\
\nonumber
&\quad +\(F^{\mu\nu}\pi\)^{I}{D}_{\mu}\pi_{I}{D}_{\lambda}\pi^{J}{D}_{\nu}{D}^{\lambda}\pi_{J}\\
&\quad +{1\over 2}\(F^{\nu\lambda}\pi\)^{I}\(F_{\mu\lambda}\pi\)^{J}{D}^{\mu}\pi_{I}{D}_{\nu}\pi_{J},\label{gaugedR}
\end{align}
whose equations for both the gauge field and scalar are second order.  Note that a specific set of non-minimal couplings has been produced.\footnote{Note that (\ref{gaugedR}) does not include the kinetic term for the gauge fields. This term would arise, along with brane Einstein-Hilbert term, from the zero mode of the bulk Einstein--Hilbert term, in a manner similar to Kaluza--Klein reductions.} This lagrangian agrees with equation (18) of \cite{Zhou:2011ix}, up to integration by parts and addition of non-minimal couplings which have second-order equations of motion.

The ansatz (\ref{gaugeonlyans}) breaks the boost symmetries of the brane into the extra dimensions, and this is reflected in the fact that the galilean symmetry of the 4D theory is spoiled by gauging.  The zero mode vectors $A_{\mu}$ and the scalars $\pi^I$ inherit a gauge transformation under the zero modes of higher-dimensional diffeomorphisms which preserve the ansatz (\ref{gaugeonlyans}), and this yields the gauge invariance of the 4D theory.

To recover the gauged and covariant galileons of \cite{Zhou:2011ix}, we turn on both the zero mode gauge fields and the zero mode metric,
\begin{align}G_{AB}&=\left (
\begin{array}{c|c}
g_{\mu\nu}(x)+A_{\mu}^{i}(x)A_{\nu}^{j}(x)\xi^{I}_i(y)\xi_{jI}(y)& A^{i}_{\mu}(x)\xi_{iI}(y)\\ \hline
A^{i}_{\mu}(x)\xi_{iI} (y)& \delta_{IJ}
\end{array}\right )\ .
\label{bulkmetric}
\end{align}
The induced metric is now given by
\begin{align}\label{gaugeandcovind}
\bar{g}_{\mu\nu}&=g_{\mu\nu}+\mathcal{D}_{\mu}\pi^{I}\mathcal{D}_{\nu}\pi_{I}\ ,
\end{align}
The covariant derivative $\mathcal{D}_\mu\equiv\nabla_\mu+A_\mu$ now acts covariantly on both gauge indices and spacetime indices.

The  calculation of $\mathcal{L}_{4}=-\sqrt{-\bar{g}}\bar{R}$ now gives, in the small field limit, the gauged and covariant galileons of \cite{Zhou:2011ix} with a specific set of non-minimal couplings, ensuring that the equations of motion for the metric, gauge fields and scalars are all second order,
\begin{align}
&\mathcal{L}_{4,{\rm gauged~ cov}}=\nn
&\quad -\mathcal{D}_{\mu}\pi^{I}\mathcal{D}_{\nu}\pi^{J}\Big(\mathcal{D}^{\nu}\mathcal{D}^{\lambda}\pi_{I} \mathcal{D}^{\mu}\mathcal{D}_{\lambda}\pi_{J}-\mathcal{D}^2\pi_{I}\mathcal{D}^{\mu}\mathcal{D}^{\nu}\pi_{J}\Big)\nn
&\quad +{1\over 2}\(F^{\mu\nu}\pi\)^{I}\mathcal{D}_{\lambda}\pi^{J}\mathcal{D}^{\lambda}\pi_{I}\mathcal{D}_{\mu}\mathcal{D}_{\nu}\pi_{J}\nn
&\quad +\(F^{\mu\nu}\pi\)^{I}\mathcal{D}_{\mu}\pi_{I}\mathcal{D}_{\lambda}\pi^{J}\mathcal{D}_{\nu}\mathcal{D}^{\lambda}\pi_{J}\nn
&\quad +{1\over 2}\(F^{\nu\lambda}\pi\)^{I}\(F_{\mu\lambda}\pi\)^{J}\mathcal{D}^{\mu}\pi_{I}\mathcal{D}_{\nu}\pi_{J}\nn
&\quad -\left (R_{\mu\nu}-\frac{1}{4}Rg_{\mu\nu}\right )\Big(\mathcal{D}^{\mu}\pi^{I}\mathcal{D}^{\nu}\pi^{J}\mathcal{D}^{\lambda}\pi_{I}\mathcal{D}_{\lambda}\pi_{J}\nn
&\quad -\frac{1}{2}\mathcal{D}^{\mu}\pi^{I}\mathcal{D}^{\nu}\pi_{I}\mathcal{D}^{\lambda}\pi^{J}\mathcal{D}_{\lambda}\pi_{J}\Big)\ .
\label{gaugedcovR}
\end{align}
This lagrangian is equivalent to equation (41) of \cite{Zhou:2011ix}, again up to integration by parts and possible addition of non-minimal couplings which retain second-order equations of motion. Note that (\ref{gaugedcovR}) is the natural fusion of (\ref{covariantizedR}) and (\ref{gaugedR}).

\section*{Conclusion}
Multi-galileon theories which are invariant under an internal global $SO(N)$ symmetry arise naturally from a co-dimension $N$ probe brane construction, in which the bulk is a fixed isotropic manifold. By allowing parts of the bulk metric to become dynamical, we have shown that the $SO(N)$ symmetry can be gauged while retaining second-order equations of motion. While we have focused on the $SO(N)$ case for concreteness, some generalization is fairly straightforward. By exploiting the embedding of $SU(N)$ into $SO(2N)$, it should be possible to couple galileons to $SU(N)$ gauge fields using the same setup with a co-dimension $2N$ bulk. Additionally, we have restricted to the case where gauge fields transform in the fundamental representation, but it should be possible to generalize to some cases of gauge fields in other representations of other groups.  The procedure would be to embed in a co-dimension $M$ bulk, such that the group $G$ we wish to represent is a subgroup of $SO(M)$, and the representation of $G$ we wish to have can be found within the restriction of the fundamental of $SO(M)$ to $G$.  Then, one would turn on only the gauge fields corresponding to $G$.

We expect that gauged galileons will have a rich and interesting phenomenology, possibly both for cosmology and for particle physics. It is possible that galileon theories may arise in beyond the standard model model physics, in particular, their non-renormalization theorem makes it very tantalizing to consider connections to long outstanding problems such as the hierarchy problem. In cosmology, galileons may arise in the dark sector. In either case, such applications will require an understand of the interplay between galileon theories and gauge fields.  It is also possible that gauged galileons may allow for interesting defect solutions, in contrast to their un-gauged counterparts \cite{Endlich:2010zj}.

{\bf Acknowledgments:} This work is supported in part by NSF grant PHY-0930521, and by Department of Energy grant DE-FG05-95ER40893-A020.

\end{document}